# Business Processes: The Theoretical Impact of Process Thinking on Information Systems Development

Mark Dumay

*Department of Information Systems & Software Engineering*, Delft University of Technology, *Delft, The Netherlands.*

## Abstract

This paper investigates two aspects of process thinking that affect the success rate of IT projects. These two aspects are the changes in the structure of organizations and the epistemology of Information Systems Development. Firstly, the conception of business processes within the management of organizations increases the structural complexity of Information Systems, because existing systems have to be integrated into a coherent cross-functional architecture. Secondly, process thinking leads to a particular view of organizations that ultimately has a negative effect on the support of Information Systems. As an illustration of process thinking, the Business Process Reengineering movement adheres to a technocratic management perspective of organizations. Particularly this conception of organization views people as mechanisms to realize certain organizational goals. As a result of this view stakeholders are confronted with the implemented systems, rather than consulted about the scope and functionality of those systems. Therefore, both aspects of process thinking have a negative impact on the success of IT projects. The problem of structural complexity is an area that is addressed by Enterprise Application Integration, and mainly requires technical solutions. However, the problems associated with the conception of organization require a different, markedly non-technical, perspective. Several directions are discussed to overcome some limitations of process thinking, but these directions are merely small pointers. If truly effective and useful Information Systems are to be acquired, IT practitioners and scientists require a completely different mindset.

## 1 Introduction

This research studies the application of Information Technology within organizations. Its related area of research – Information Systems Development – is dominated by process thinking. The effects of process thinking on the success rate of IT projects are unknown. To investigate these effects, this section elaborates on the setup of research.

### 1.1 Research Background

Annual expenditures on Information Technology by organizations in the United States are estimated at $250 billion. According to the Standish Group (2003) 51% of all IT projects are challenged, meaning that those projects are either not finished on time or fail to remain within their budget constraints. In practice these conditions often occur simultaneously. The average budget overrun is estimated at 43%, while the deadline is likely to be exceeded by 82% of the original project's allocated time. Another 15% of the projects are entirely abandoned, which leaves an overall success rate of 34% for IT projects. Although these figures may not be directly applicable to the situation in the Netherlands, they do give an impression of the still

eminent struggle of IT specialists to implement technology at organizations in a predictable and manageable way.

The development of organizational management in the last decade could be a clue to the low success rate of IT projects. During the 1990's, many organizations restructured their departments when they applied a process perspective to their business. This process perspective superseded their earlier functional one (Silvestro and Westley 2002: 217). Inspired by the success of Japanese car manufacturers in the 1970's and 80's[1], Harrington (1991) devised a strategy for Total Quality Management (TQM)[2]. Others described an even more disruptive intervention known as Business Process Reengineering (BPR) (Scott Morton 1991; Davenport 1993; Hammer and Champy 1993). Both of these movements share a focus on business processes as a key indicator for the ability of an organization to deliver its products or services in a timely and efficient manner (Hammer 1996: 83). Such a fundamental shift in the perspective of management likely has large implications for existing IT infrastructures, and complicates the development of new Information Systems.

Another clue to the failure of Information Technology within organizations is the social impact of such technology. Even as early as 1988 – just before the emergence of BPR within management journals – it was suggested that IT projects are not likely to fail due to technical complications, but seem to suffer from contradicting interests of the various stakeholders (Riesewijk and Warmerdam 1988). Perhaps the complicated nature of IT implementations are not merely the result of the aforementioned restructuring of organizations. Both the TQM and BPR movement conceive organizations using a perspective that can be labeled as 'process thinking'. Its supportive technique is Business Process Modeling, which is dominant for IT consultants trying to define the organizational context of Information Technology (Butler, Bahrami et al. 2000). The area of research that studies the relationship between Information Technology and the organization as context is called Information Systems Development. The dominance of process thinking within this scientific discipline could be another cause of the high failure rates of IT projects.

## 1.2   Research Questions

As mentioned in the previous section, process thinking is claimed to have at least two major implications on the success rate of IT projects. Firstly, the restructuring of organizations by focusing on business processes – instead of tasks and functions – has consequences for existing and future Information Systems from a structural point of view. Secondly, it influences the consideration of the social impact of Information Systems during the development and implementation of such systems. The aim of this paper is to investigate these claims, and to give directions on how to overcome their resulting limitations. Therefore, this paper deals with the following questions:

1. Does the impact of business processes on the structure of organizations require a different practice within the area of Information Systems Development?
2. Is the conception of organization present in process thinking adequate to define the context of Information Systems?
3. What are the limitations of process thinking within Information Systems Development and how can they be overcome?

The first question encompasses both the emergence and definition of process thinking. The application of this perspective to organizations gives insight into the structural aspect of Information Technology within organizations. Secondly, the consideration of the social-organizational context of such technology is dependent on the applied view of organizations – the conception of organization. Finally, the limitations of process thinking are to be investigated. Should these limitations exist, this paper will attempt to provide directions in overcoming them.

---

[1] Jaffee (2001: 132) gives a historical overview of the production model of Japanese car manufacturers, who adapted and improved on the traditional Ford assembly line.
[2] In the Netherlands the 'Model Nederlandse Kwaliteit' (trans. Dutch Quality Model) was a notable set of TQM guidelines for the Dutch industry (Smit 2003). Nowadays, this model is known as the 'INK-managementmodel' (trans. INK Management Model) (INK 2003). The EFQM Excellence Model is the leading organizational framework in Europe, partially based on TQM principles (EFQM 2003).

## 1.3 Epistemological Justification of Conducted Research

In my opinion the study of organizations cannot be defined as an exact science. Organizations evolve in the interaction of participants; an evolution that is dependent on the shared convictions and values of its participants. I also reject the idea that a single, universally appropriate type of organization exists, or that a single utopian view on organizations can be applied to each and every organization in real life. These notions may conflict with my background in Computer Science, a field that is often regarded as 'hard' science. I do think, however, that the application of Information Technology within organizations justifies a broader perspective. The social scientific movement that resembles my ideas is social constructivism (Kroes 1996). This movement rejects the idea of separating knowledge from its social background, thereby implying that knowledge is subjective (Jones 2002). Social constructivism also states that there is no single objective reality, since reality is constructed by the process of reaching a shared universe of discourse between the various participants. Social ontological facts are therefore negotiated, which can be viewed as ontological relativism. One could describe this relativism as an extreme or radical form of social constructivism.

Most and for all I apply social constructivism on a meta-theoretical level of social scientific research. Within the social sciences I observe different schools of thought are separated by their respective ideology. Each of these schools has its own accepted research methodology and practice, which I consider to be a subject of negotiation among the school's participants. In addition, I presume these schools to coexist and complement each other in their contribution to social scientific research in general. This presumption implies a single ontological view on organizations is not possible, since each conception of an organization is bound to its respective ideology. Burrell and Morgan (1979) conducted research on organizational analysis and identified four basic paradigms. Their definition of a paradigm equals my view on schools of thought and their ideologies: '[a paradigm is] a term which is intended to emphasize the commonality of perspective which binds the work of a group of theorists together in such a way that they can be usefully regarded as approaching social theory within the bounds of the same problematic' (Burrell and Morgan 1979: 23). A critical difference of this view with the paradigm definition of Kuhn, is that Kuhn described the historically *successive* developments of the natural sciences, whereas Burrell and Morgan assume theories in the social sciences to *coexist* (Iivari, Hirschheim et al. 1998: 171-172).

On the level of applied research, I do think social constructivism – which I relate to the interpretive paradigm of Burrell and Morgan – is an important contribution to the understanding of organizations. However, I would like to stress that I do not wish to apply social constructivism in an imperialistic way. I think the real strength of social constructivism is its acceptance of different ideologies. Furthermore, since most practitioners in the field of Information Systems are guided by the philosophical assumptions of functionalism (Goles and Hirschheim 2000), they are not likely to accept an interpretive critique of their practice. To overcome this ideological difference, I will use an epistemology known as paradigm interplay. This epistemology allows for acknowledging both differences and similarities of the functionalist and interpretive paradigms (Goles and Hirschheim 2000: 260), while also bridging ideological differences between their respective followers. Within the functionalist tradition, I will reevaluate the conception of organization using a modernist epistemology; this is known as post-modern periodization[3]. This conception will be used to discuss the impact of Business Process Reengineering – an illustrative movement that adheres to process thinking – on both organizations and Information Systems Development. Critical identified aspects will be reflected from an interpretive point of view. A more precise description can be found in the next section.

---

[3] Parker (1992) discusses two distinct positions within postmodernism, which he describes as 'postmodern epistemology' and 'post-modern periodization' (the hyphenation is deliberate). While the former movement completely rejects modernism and realism, the latter simply reapplies modernist approaches to an observable empirical reality. Both agree that new forms of organization are emerging, but differ in their ontological and epistemological assumptions on how to comprehend this development.

## 1.4   Outline of This Paper

The questions discussed in §1.2 impose the structure of this paper. Respectively chapter 2, 3, and 4 address these three questions. Chapter 1 is an introduction to the background of the research (§1.1), and describes the related research questions (§1.2). The research methodology to obtain the answers to these questions is based upon an epistemology of research (§1.3). These answers form the outline of this paper (§1.4).

Chapter 2 illustrates the concept of process thinking and how this concept affects the structure of organizations and their IT systems. The emergence of Total Quality Management and Business Process Reengineering illustrates the context in which process thinking surfaced (§2.1). Business Process Reengineering is chosen as an example of how such a perspective can be used and implemented in organizations (§2.2). This implementation has several effects on Information Systems within organizations (§2.3). The conclusion answers the question if the current practice of Information Systems Development is sufficient to deal with the identified effects on Information Systems (§2.4).

Chapter 3 studies the conception of organization according to process thinking. A paradigmatic view of organization provides the necessary broader perspective to investigate this conception (§3.1). General Systems Theory serves as a bridge between the paradigmatic theory and the conception, because General Systems Theory is widely used within the paradigm to which process thinking belongs (§3.2). An analysis of Business Process Reengineering illustrates the conception of organization according to process thinking (§3.3). The conclusion denominates the deduced conception and elaborates on its conflict with the General Systems Theory (§3.4).

Chapter 4 reflects on the research findings and identifies several limitations of process thinking. Critical analysis of Business Process Reengineering is a starting point to obtain these limitations (§4.1). Several directions to overcome some of these limitations with respect to Information Systems Development are presented (§4.2). The conclusion elaborates their practical use (§4.3). Finally, chapter 5 gives the overall conclusion of this research and stresses the limitations of the research conducted.

## 2   The Organizational Shift from Tasks to Processes

The development of organizational management in the early 1990s sheds light on the context of process thinking as meant in this paper. To illustrate the application of process thinking to the restructuring of organizations, this section deals with the orientation's most extreme movement: Business Process Reengineering. The reforms initiated by BPR have had several implications on the IT infrastructure of these organizations.

## 2.1   Emergence of Process Thinking

In 1990, two articles by Hammer (1990) and Davenport and Short (1990) created quite a stir within both theory and practice of organizations (Barothy, Peterhans et al. 1995). Earlier on, in 1984, Scott Morton devised a research program – The Management in the 1990s Research Program – to explore the impact of Information Technology on modern organizations (Scott Morton 1991: 3). In the same year (1991) of Scott Morton's publication of the program's results, Keen published a book with a similar subject. While the latter was rather vague in describing his orientation – a team-based collaborative organization (Keen 1991: 108) – Hammer coined the term Business Process Reengineering (Hammer 1990). Davenport and Short (1990) labeled their movement Business Process Redesign, which was later superseded by the term Business Process Innovation (Davenport 1993). Although a lot of different names are present, they all represent a movement that suggests organizations need to radically transform their current practice. Only then will they be able to cope with the high demands of the business environment.

Business Process Reengineering (BPR) – which is the label used throughout the remainder of this text – is complementary to another movement. This movement – known as Total Quality Management (TQM) (Harrington 1991) or 'kaizen' (Davenport 1993: 312) – shares the process view of organizations with BPR. Instead of considering structure and control – illustrated by the administrative organization of Simon (1976) – both the TQM and BPR process orientations focus on overall performance from a client perspective. Within that context, existing functional divisions are likely to hinder the throughput of products and services, since each hand-off between departments creates extra delay. In addition, errors are more likely to occur due to miscommunication. The term 'reengineering' should be understood in the context that Michael Hammer used it in initially. In his famous article of 1990, he used the Ford Motor Company as an example for the entire American business world. Beaten by their Japanese competitors in quality, price, and time-to-market in the 1970s and 1980s, Ford radically overhauled its organization. The company cut many old bureaucratic habits and focused on the overall performance of its processes instead. Hammer used the term reengineering to stress that the entire company needed to change radically, and should not settle with minor improvements.

Davenport analyzed the key differences between TQM and BPR (Malhotra 1998). TQM can also be regarded as a continuous improvement program, since quality management is an ongoing process. BPR is more concerned with radical change, which is typically linked with the concept of innovation. The differences are represented in Table 2-1.

|  | Improvement (TQM) | Innovation (BPR) |
|---|---|---|
| Level of change | Incremental | Radical |
| Starting point | Existing process | Clean slate |
| Frequency of change | One-time/continuous | One-time |
| Time required | Short | Long (at least 2 years) |
| Participation | Bottom-up | Top-down |
| Typical scope | Narrow, within functions | Broad, cross-functional |
| Associated risk | Moderate | High |
| Primary Enabler | Statistical control | Information technology |
| Type of change | Cultural | Cultural/structural |

**Table 2-1: Differences between TQM and BPR (adapted from Davenport 1993: 11)**

If the Ford case is taken into account, BPR was originally envisioned as a one-of-a-kind change program. This is also present in the overview of Davenport. Until the late 1980s Western business companies[4] were inefficient at realizing their goals, because they were slow, bureaucratic organizations. Many firms used the *principle of the division of labor* as their primary structural management methodology (Hammer and Champy 1993: 12). Adam Smith, who wrote his famous book 'The Wealth of Nations' in 1776, argued that – in order to enable mass production – labor should be divided into many specialist tasks. Alfred Sloan applied this orientation to management, by introducing divisions and hierarchical structures. In practice, this means that growing organizations require more and more overhead, since the span of control – the amount of people one manager can supervise – is limited. The increasing distance between senior management and the shop floor hinders communication. The end result for many companies was that they were unable to timely respond to the changing demands of customers. BPR was therefore a radical change program, because it aimed to abandon the dominant command-and-control system present in many organizations. The favorable type of organization was the 'process organization'.

Next to business processes – a concept borrowed from TQM – and the vision of radical change, BPR has another key ingredient. The two most influential names in the BPR movement are associated with Information Technology. Michael Hammer is a former computer science professor at MIT (Hammer and Champy 1993), while Thomas Davenport was a partner at Ernst

---

[4] Note that both Davenport and Hammer really meant *American* companies, when they talked about Western businesses. The understanding and implementation of BPR within Europe has been much more diffused (Newell, Swan et al. 1998). The historical overview in this section is based on Hammer and Champy (1993), and therefore does not necessarily reflect the European situation.

& Young's Center for Information Technology and Strategy in Boston (Davenport 1993). Both authors worked together in a multiclient research program called 'Partnership for Research in Information Systems Management' (PRISM) in 1988 (Davenport 1995). An important occasion of this program and other related programs is the productivity paradox[5]. This paradox is an observation about the effect of IT on the productivity of companies in the USA. Despite the increasing investment in computing power from the 1970s and onwards, the productivity – especially of the service sector – stagnated (Brynjolfsson 1993; Petrovic 1995). Whether this paradox was true or not[6], it was a major argument for many practitioners and researchers who stressed that IT was applied wrongly to organizations. It comes as no surprise, then, that the role of IT in BPR is eminent.

Although radical change, business processes, and Information Technology remain the key ingredients of BPR, the meaning and implication of BPR changed over the 1990s. For instance, Hammer (1996) stressed that the key concept of BPR was no longer the word 'radical', but just 'business process'. One can argue that improvement and innovation programs are still necessary, even when the organization has left its functional divisional structure behind. In addition, the distinction between the TQM and BPR approaches is clear in theory, but the dividing line is often blurred in practice (Altinkemer, Chaturvedi et al. 1998:381-2). BPR just becomes part of the management armamentarium. The renewed relationship between TQM and BPR is illustrated in Figure 2-1. The quality programs are equal to the concept of continuous improvement, and reengineering is similar to Davenport's notion of innovation. Most of Davenport's earlier observations about BPR remain valid, though. BPR is more radical than TQM, because it is less bound to existing work arrangements. Innovation should not be limited by the implementation of current business processes at first, because this will likely narrow the area of study too much. BPR is therefore exemplary as an illustration of process thinking. It is a clean break with organizations of the past, which were dominated by functions, tasks, and structures.

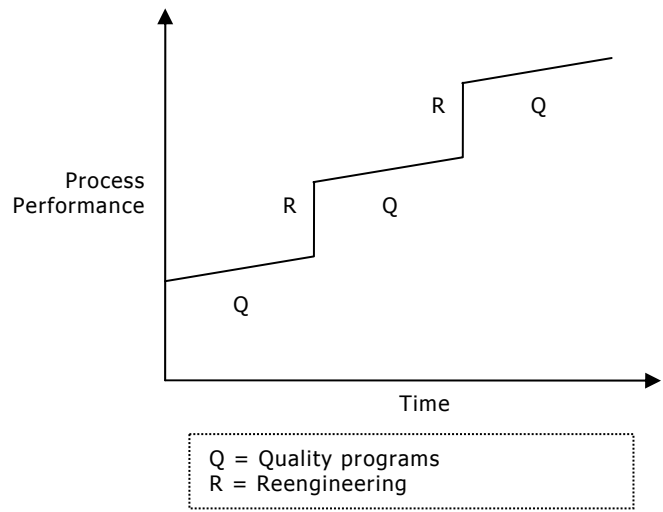

**Figure 2-1: Relationship of TQM and BPR throughout time
(adapted from Hammer 1996: 83)**

---

[5] References to this paradox are present explicitly in Scott Morton (1991:v-vii) and Davenport and Short (1990: 12). Others simply refer to the required increase of awareness of IT among managers (cf. Keen 1991).
[6] For example, Greenstein (1996) questions if the productivity really stagnated and therefore rejects the entire paradox.

## 2.2   The Implementation of BPR within Organizations

The previous section discussed the context in which Business Process Reengineering (BPR) emerged. The BPR approach has its roots in IT management[7] (Malhotra 1998), and was only later addressed by science. Especially during the period 1990-95 BPR reached a peak in terms of management hype (Davenport 1995), while at the end of that period it was just gaining momentum within scientific research (Barothy, Peterhans et al. 1995). Also it was mentioned that BPR has become part of management practice, in addition to approaches such as Total Quality Management (TQM). To understand the original envisioning of BPR, the steps involved in implementing BPR as was common during the early 1990s are discussed.

Barothy called for the foundation of BPR as a scientific research subject. His review of publications prior to his article – including the books from Hammer and Champy and Davenport – leads to the following definition of BPR: '*a complex, top-down driven and planned organizational change task aiming to achieve radical performance improvements in one or several cross-functional, inter- or intra-organizational business processes whereby IT is deployed to enable the new business process(es)*' (Barothy, Peterhans et al. 1995: 4). This definition reflects the context in which BPR emerged, as discussed in the previous section. A high-level approach to implement BPR within organizations is given by Davenport (1993: 25). A more detailed discussion is given by Schwartz, Hwang et al. (1995). The five major steps of the BPR methodology are summarized as follows.
1. Identify processes.
2. Identify change levers.
3. Develop process visions.
4. Understand existing processes.
5. Design and prototype the new process.

The first step is in effect a critical assessment of the organization's strengths and weaknesses. The major business processes are enumerated and their viability is judged. In this phase it is not necessary to know the complete details of each and every business process, since the main focus should be on overall performance and strategic relevance. Next, it is important to identify the potential opportunities – especially technological and human openings. Likewise, possible constraints have to be considered as well. The third step involves the necessary vision of the business processes that are selected for reengineering. Especially performance targets and benchmark results can contribute to the rationale of the change program. If the vision is formulated, the present implementation of the business processes should be understood. If tested along the objectives, shortcomings and problems can be identified more easily. Finally, a brainstorm session to obtain design alternatives should lead to the selection of a best alternative and the subsequent development of a prototype. Implementing this new design requires a proper migration strategy. As is apparent in this approach, BPR is envisioned as a top-down exercise. Barothy's definition emphasizes the importance of IT within a BPR change program, whereas Davenport is keen to stress the importance of human change levers as well.

The role of Information Technology within BPR is eminent: not only is IT a facilitator for BPR implementations, but it is also an enabler of organizational change. Hammer and Champy (1993: 84) label this as *inductive thinking*: 'the ability to first recognize a powerful solution and then seek the problems it might solve'. New technologies rarely have a direct application, but if a function is found it will give a striking competitive advantage. Davenport (1993: 200) adds that IT can be supportive *during* BPR change programs as well. This varies from tools to support the design of new business processes, performance monitoring tools, and collaborative technology such as e-mail and directory services.

After the business processes are designed and prototyped, they are to be embedded permanently within the organization. The necessary change from a functional structure towards a process-based organizational structure is the greatest challenge. Although some degree of formal structure remains necessary, work is no longer built around specific skills and tasks (Davenport 1993: 161). Both Hammer and Champy (1993) and Davenport (1993) introduce the concept of the case worker and the case team. Such a person or team is responsible for an entire customer deliverable. This empowerment gives employees the authority to make decisions required to solve customer issues instantly, without having to wait for management's approval. In addition, role expansion is necessary, because several functions that were split across departments are now combined within the responsibility of one person or team. Both empowerment and role expansion are a direct result of the client orientation of business processes in BPR (see §2.2), because they shorten processing time. The assumption of BPR is that the problems of handoff between departments are hereby eliminated.

---

[7] The two most influential articles of Davenport and Short (1990) and Hammer (1990) were both published in management journals. Davenport (1995: 70) stresses that the 'real' creators of reengineering are managers and other practitioners. Section 2.1 discusses the role of IT.

## 2.3   Effect on Information Systems within Organizations

The use of Information Technology (IT) in organizations is typically linked with the concept of *Information Systems*. Without further defining this concept here, its mission as proposed by McNurlin and Sprague (1989: 12) will shed some light on the role of IT within organizations: 'To improve the performance of people in organizations through the use of information technology'. Apparently the use of IT within organizations should support a certain goal or purpose, at least from a management perspective[8]. This implies that software and hardware are adapted in a wider context – the organization or a subunit – and their use is – or should be – well-considered.

Especially during the 1980s, the awareness of the strategic value of Information Systems increased (McNurlin and Sprague 1989: 58-86). Daft (1998) identifies several types of Information Systems. Their application has shifted from 'efficient machines' towards 'strategic arms' in the last couple of decades. Transaction Processing Systems are an illustration of a system with a low level of complexity. Such a system could be an airline reservation system, or an ordering system. These types of system are typically associated with lower levels of management, and therefore do not have much impact on the organization's strategy. Management Information Systems and Decision Support Systems form a different category, since they are more coupled with the decision-making within organizations. Such systems have an increased level of complexity, as opposed to efficient machines. The final type of Information System can be labeled as strategic arms, and this coincides with the observation of McNurlin and Sprague (1989) about the increased strategic value of Information Systems in the 1980s. Daft (1998) places Executive-management Information Systems, Networks, and Electronic Data Interchange in this category. The development of Information Systems is visualized in Figure 2-2.

---

[8] Barbara McNurlin and Ralph Sprague had a management perspective when they discussed the use of IT within organizations. This perspective – which is also present in the work of Richard Daft (1998) – is maintained throughout this paragraph.

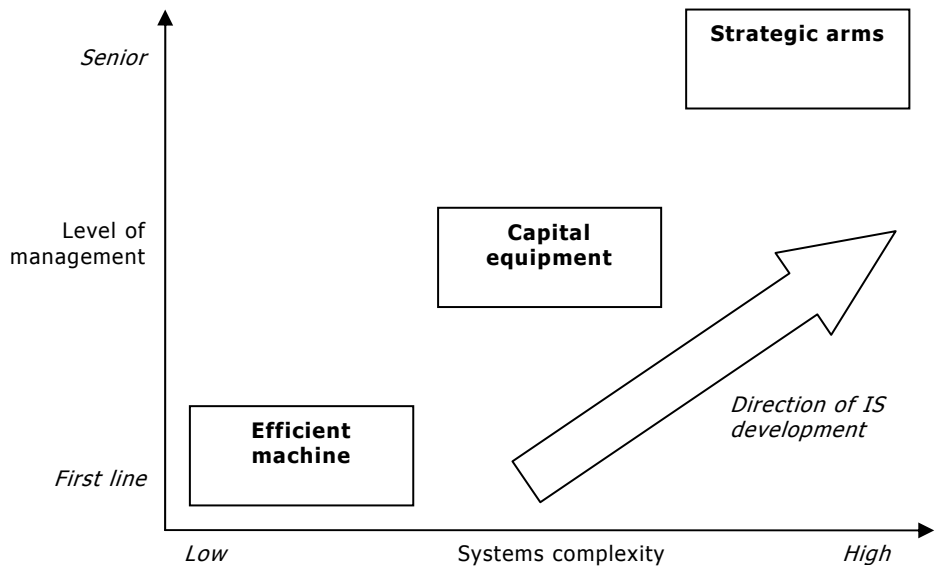

**Figure 2-2: Development of IS within organizations (adapted from Daft 1998: 280)**

The application of the process orientation to organizations (see §2.1 for an elaboration) has several implications on existing Information Systems, but also on the development of new systems. The most fundamental characteristic of business processes with respect to Information Systems is that they are cross-functional. In the past, Information Systems were bound to a functional division or a subunit of the organization. If these borders are surpassed, the existing systems have to be integrated[9]. This is known as Enterprise Application Integration (Johannesson and Perjons 2001: 165). While this is largely a technical exercise – concerned with techniques such as middleware – it has great impact on the management of these systems. As can be seen in Daft's development of Information Systems, the complexity of these systems increases dramatically. Melling (1994: 496) observes that Transaction Processing Systems (categorized by Daft as 'efficient machines') are typically task oriented. This suggests that the highest level of systems complexity deals with cross-functional business processes, since they are the highest abstraction level of the production or service provision of the organization.

The alignment of Information Technology and business strategy is a topic that can be related directly to Daft's notion of the development of Information Systems within organizations. If the strategy of the organization is leading, and management realizes the strategic value of Information Technology, the IT Architecture can be seen as the realization of the business strategy (Earl 1989: 135). Such architecture consists of elements as applications, networks, and standards. Melling (1994: 495) stresses that BPR 'is a key element in the factors driving IT architectural change'. If the business operations of an organization change radically, the IT framework will have to change accordingly. In this respect, Enterprise Application Integration can be put in a wider context provided by the concept of the IT Architecture. Nowadays, this topic is still being addressed by many leading (Dutch) software vendors, like IBM, Capgemini, and Ordina.

## 2.4 Conclusions

Both Business Process Reengineering (BPR) and Total Quality Management (TQM) focus on business processes within organizations. The business process represents the means to deliver a product or service, whereby the needs of the customer are of primary concern. It is believed that organizations until the late 1980s are predominantly concerned with formal structures, functions, and control mechanisms. This leads to many mistakes due to handoffs between departments, and long delays because of authorization issues. By focusing on business processes instead, these issues can be solved. This requires that organizations have to be *reengineered*. BPR is the most extreme illustration of process thinking, as it seeks to replace the entire organizational structure with core business processes that are supported by case teams. These teams are largely self-maintaining and thus require little supervision. Therefore the middle management of organizations is superfluous.

---

[9] In most cases it is simply not feasible to replace the entire IT infrastructure and systems with a better designed one.

The strategic importance of Information Technology increased in the years prior to BPR, and continues to do so nowadays. IT is a key issue of BPR, as it is both an important enabler and facilitator of business processes. Because business processes are cross-functional and exceed boundaries between departments, existing Information Systems have to be integrated. In addition, BPR strongly advocates the strategic importance of Information Technology. Both aspects contribute to an increased level of complexity of Information Systems, since the scope of application of the technology is enlarged. However, the increased level of complexity is mainly a technical issue, which Enterprise Application Integration addresses. Therefore, Information Systems Development does not need to change its epistemology to cope with the aspect of structural complexity.

## 3   The Conception of Organization within Process Thinking

To understand the organizational orientation of process thinking, a wider context of organizations is required. This context is provided by a paradigmatic view of sociology. Within this view, the paradigm that best-fits process thinking is discussed. In addition, the General Systems Theory is explained, as it functions as a bridge between that paradigm and the application of process thinking. Finally the conception of organization according to Business Process Reengineering – chosen as an example of process thinking in the previous chapter – is elaborated on in terms of a model of organization.

### 3.1   A Paradigmatic View on Organization

Organizations are studied by numerous practitioners and scientists (Jaffee 2001: 1). Unlike natural phenomena, such as biological cells and the solar system, organizations are a human conception. Simply to view organizations as a collection of people does not do justice to the impact of organizations on society. Since the study of organizations is such a diverse and multidisciplinary field of research, one can presume the organization as a concept is quite intangible. To cope with the complexity of the phenomenon, many researchers have a specific frame of reference when studying organizations. For example, economists are likely to view organizations as the means to realize an objective of profit[10]. Psychologists focus on the individuals within organizations, and give less attention to group behavior or the organization's environment. Sociology primarily studies the collective behavior of individuals. As Jaffee (2001: 208) points out, the contribution of sociology to science in general is the insight that individuals operate in a social environment. Although many question the value of sociology nowadays (Casey 2002), it does encompass a level of analysis that is deemed most appropriate for this research.

The term 'sociology' was first coined between the social philosophers Saint-Simon (1760-1825) and August Comte (1798-1857).They endeavored after the establishment of a 'science of social physics' (Casey 2002: 30). This mention of 'physics' implies that they envisioned a rational and structured research methodology to study the fundamental concepts within this emerging field of research. This positivist epistemology has been dominant until the mid-20$^{th}$ century (Casey 2002: 88), when counter movements battled this dogma in what has been labeled the 'paradigm wars' (Goles and Hirschheim 2000). The paradigm wars identified a critical division between the fundamental assumptions of social scientific researchers, which could be separated into objectivists and subjectivists (Burrell and Morgan 1979). The objectivists – in tradition with the dominant positivist epistemology – supported ontological realism, whereas the subjectivists adhered to ontological nominalism. An objectivist claims there is a single objective reality that can be identified by a thorough research methodology. Subjectivists on the other hand question the existence of such a reality; let alone that this reality could be described independently from its observers. For subjectivists, a phenomenon is only labeled out of convenience, and this label should not be mistaken for a universal truth.

---

[10] The descriptions non-profit organizations or not-for-profit organizations suggest economists are open to other organizational goals. At the same time, these descriptions still illustrate the focus on monetary results.

The debate between objectivists and subjectivists has not been part of sociology alone. For example, in the field of genetics the work of Barbara McClintock was questioned by many fellow researchers during the 1940s and 50s (Comfort 1995) and did not receive any real credit until the late 1960s and 70s[11]. McClintock experimented with corn plants in a very labor-intensive way. Instead of treating those plants as uniform crops, she identified herself with each and every single plant. The research she presented was a welter of data, where each and every important mutation was discussed in light of an extensive and detailed background description (Comfort 1995: 1165). In retrospect, McClintock did not rely on ontological realism, but adapted an epistemology rooted in ontological nominalism. Later on, in 1983, she received a Nobel Prize for her discovery of 'jumping genes'.

In addition to the aforementioned philosophical conflict among sociologists, Burrell and Morgan (1979) identified a conflict in the sociologist's theory about the nature of society. The *sociology of regulation* emphasizes the underlying unity and cohesiveness of society, while the *sociology of radical change* tries to explain the unstable nature of society. Burrell and Morgan state that 'all theories of [organization] are based upon a philosophy of science and a theory of society' (Burrell and Morgan 1979: 1). The identified conflicts correspond with the two distinct parts of their theory of organizations. These conflicts shape the two dimensions within a matrix that is used as a typology to identify the basic paradigms[12] within sociology. These paradigms are visualized in Figure 3-1.

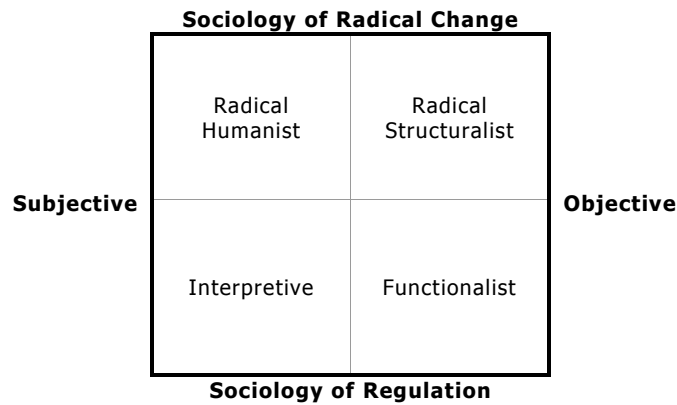

Figure 3-1: Paradigms of social theory (adapted from Burrell and Morgan 1979: 22)

The functionalist paradigm and the interpretive paradigm share the basic assumption that society tends to be orderly and stable. Such consensus may be interrupted by temporary disagreement, but it is not overruled. But where functionalists assume this reality is ontologically prior to man, interpretivists stress that consent is the result of discourse between participants. Another dispute is the position of the observer. Within the functionalist paradigm, the observer can ultimately describe reality by rational means. This is contrary to the viewpoint of interpretivists, who object to the idea of an observer who does not interfere with the situation he tries to understand (the so-called Hawthorne effect[13]). Especially within sociological studies the participants are likely to interact with the researchers in some way – be it intentional or unintentional – and alter their behavior accordingly[14]. In addition interpretivists question the linear relationship of cause and effect implied by rationality. In short, functionalists tend to use nomothetic approaches, whereas interpretivists adhere to idiographic approaches[15]. If radical change is favored over regulation, the paradigms of the radical humanists and the radical structuralists come into play. Supporters of the former focus on all forces and barriers within society that destabilize social institutions, in order to understand and overcome them. On the other hand, radical structuralists emphasize 'the need to overthrow or transcend the limitations placed on existing social and organizational arrangements' (Goles and Hirschheim 2000: 254).

---

[11] Source: the historical description of Barbara McClintock in the *Encyclopædia Britannica (2004)*.
[12] See §1.3 for a comparison with the paradigm as described by Kuhn.

## 3.2   General Systems Theory Applied to Organization

The previous paragraph mentioned that the functionalist paradigm has played an important – at times even dominant – role in the field of sociology throughout the 20$^{th}$ century[16]. Within that paradigm – which can be identified in other fields of research as well[17] – the work of Von Bertalanffy has been very influential. As the founder of the General Systems Theory (Von Bertalanffy 1950; Boulding 1956), his view of systems has had a lot of impact on many scientific disciplines. He responded to the reductionist approach of science – which had been dominant in the early 1950s – by propagating a holistic perceptive view. Reductionism highlights a single element or an aspect of a (complex) object of study, which hinders the later integration of the various findings in a single, coherent description (Nadler 1995: 129). Von Bertalanffy was one of the scientists to notice the interdependence of these elements, and suggested that many empirical properties of an object of study can not be reduced to a single element (Van Dinten 2002: 92). Von Bertalanffy's goal of his systems approach was to demonstrate that most scientific disciplines were related to each other in an abstract and general way (Scott 1992: 76). Entities such as a biological cell, a group of people, or our solar system can all be regarded as a system. Although the object of study can be quite different among various disciplines, the scientific approach to obtain knowledge about the object is – or should be – largely homogeneous.

A very basic definition of a system is that of *a whole consisting of interrelated parts* (Morin 1992). Although systems are widely accepted as an analytical tool to study a phenomenon, they remain conceptual constructions, superimposed on the philosophical concept of *object*[18]. Firstly, the definition of the system's boundary is quite arbitrary (Flood and Carson 1988: 7). For instance, it can be useful to enlarge these boundaries if there is an intense interaction between an element and its environment. Secondly, a system requires an explicit description of its elements. Even though Von Bertalanffy strongly advises that these elements should be studied in a coherent manner, one can assume he regarded these elements to be distinguishable from their context. When the systems concept is applied to different areas of research, the fluctuating level of complexity of the imposed system is eminent. Boulding (1956) is one of the researchers who differentiated the complexity of various systems. He described nine different levels of increasing complexity. Every type of system on a succeeding level inherits the complexity of its predecessor, while adding its own. An overview of these system levels is presented in Table 3-1. Although this typology seems intuitively appropriate, it does not provide a definition or even understanding of the gaps between the various levels (Checkland 1991: 106).

---

[13] Jaffee (2001: 65-73) gives a historical overview of the experiments conducted at the Western Electric Company plant in Hawthorne. In addition, he gives a thorough analysis on the consequences of the research project on the field of organization theory.

[14] A particular research methodology that builds upon the idea that the researcher should be part of the object of study – in order to gain better understanding – is 'action research'. Kock (2004) gives an overview of the application of this methodology within the field of Information Systems Development.

[15] Brown (1992: 154-5) describes these two types of approaches as "the study of a single variable in many subjects for the purpose of discovering general laws or principles of behavior" and "the thorough study of individual cases, with emphasis on each subject's characteristic traits" respectively. I would like to stress that generalizing outcomes of idiographic studies to a general model is not without severe limitations. Such an application is rejected from an objectivist point of view, because it has low quantitative backing. In addition, subjectivists should be aware that models are limiting when perceiving or understanding a phenomenon. Van Dinten (2002) illustrates these limitations with an analysis of Checkland's Soft Systems Methodology and Luhmann's autopoietic system.

[16] Parker (1992) stresses that versions of functionalism remained dominant after the mid-20$^{th}$ century, at least until the early 1990s. Apparently the paradigm wars mentioned in the previous paragraph did not diminish this dominance too much.

[17] For example, Mingers and Brocklesby (1997) identify three different paradigms: empirical-analytic, interpretive, and critical. The empirical-analytic paradigm is synonymous with the functionalist paradigm of Burrell and Morgan.

[18] The word object here is meant as *object of study*. Although this suggests such an object could be clearly defined and be isolated from its environment, I would like to stress that such a separation would be implied by its observer, and is not imposed naturally or physically per se.

| Level | Characteristics | Examples | Relevant disciplines |
|---|---|---|---|
| 1. Structures, Frameworks | Static | Crystal structures, bridges | Description, verbal or pictorial, in any discipline |
| 2. Clock-works | Predetermined motion (may exhibit equilibrium) | Clocks, machines, the solar system | Physics, classical natural science |
| 3. Control mechanisms | Closed-loop control | Thermostats, homeostasis mechanisms in organisms | Control theory, cybernetics |
| 4. Open systems | Structurally self-maintaining | Flames, biological cells | Theory of metabolism (information theory) |
| 5. Lower organisms | Organized whole with functional parts, 'blue-printed' growth, reproduction | Plants | Botany |
| 6. Animals | A brain to guide total behavior, ability to learn | Birds and beasts | Zoology |
| 7. Man | Self-consciousness, knowledge of knowledge, symbolic language | Human beings | Biology, Psychology |
| 8. Socio-cultural systems | Roles, communication, transmission of values | Families, the Boy Scouts, drinking clubs, nations | History, sociology, anthropology, behavioral science |
| 9. Transcendental systems | 'Inescapable unknowables' | The idea of God | ? |

**Table 3-1: Boulding's hierarchy of real-world complexity (adapted from Checkland 1991: 105)**

When an organization is viewed as a system, several common attributes can be identified. Burrell and Morgan (1979: 63) describe the following general principles of systems theorists, when they apply their theory to organizations:

a) that the system can be identified by some sort of *boundary* which differentiates if from its environment;
b) that the system is essentially *processual* in nature;
c) that this process can be [conceptualized] in terms of a basic model which focuses upon *input*, *throughput*, *output*, and *feedback*;
d) that the overall operation of the system can be understood in terms of the satisfaction of the *system needs* geared to survival or the achievement of *homeostasis*;
e) that the system is composed of *subsystems* which contribute to the satisfaction of the system's overall needs;
f) that these subsystems, which themselves have identifiable boundaries, are in a state of mutual *interdependence*, both internally and in relation to their environment;
g) that the operation of the system can be observed in terms of the [*behavior*] of its constituent elements;
h) that the critical activities within the context of system operation are those which involve *boundary transactions*, both internally between subsystems and externally in relation to the environment.

The concept of systems is rooted in biology; a scientific discipline in which Von Bertalanffy elaborated his theory. In such an area of research, elements can be related to physical, empirical entities with relative ease. But if the system's theory is applied to organizations, the single most complicating factor is the enormous complexity of the phenomenon. Organizations are socio-cultural systems within Boulding's hierarchy, and such systems have the highest level of complexity, next to transcendental systems. This makes it hard or impossible to disseminate the system's elements in an unbiased way. In addition, Burrell and Morgan (1979: 68) conclude that most social theorists reach for some kind of analogy in advance of any system to which it is applied. Such an analogy – that is likely to be mechanical or organismic – reduces the complexity of the system to an almost elementary model. This is a common problem within systems theory (Morin 1992: 372), but it is especially present in the study of organizations, due to the paramount complexity of organizations compared to other phenomena (if Boulding's hierarchy is taken into account).

The problem of mapping real-world phenomena to system elements has been addressed by several researchers. Peter Checkland, a renowned researcher in the field of Problem Structuring Methodology (Jackson 1991; Mingers 2000; Van Dinten 2002), distinguishes between soft systems and hard systems in his soft systems methodology. He adds on Von Bertalanffy's definition of a system by including the perception of the system's designer; something he describes as systems thinking (Checkland 1991: 102). Hard problem situations clearly and explicitly define the problem at hand and the elements involved. Within their definition, goals are known, and the performance of possible solutions is measurable. However, soft problems – which include organizational problems – possess quite opposite properties. In this respect the definition of a system should be the end result of a participative discussion between all stakeholders involved. Checkland's methodology aims at a structured discussion about conceptual models – each highlighting a specific aspect of the problem situation – to reach an agreement between those involved (Jackson 1991: 156). Checkland's solution to the mapping problem therefore lies in consensus: as long as all participants agree on the systems description of the object under debate, such a system can be regarded as 'valid'.

### 3.3   The Conception of Organization According to BPR

As is argued in §3.2, the primary problem when viewing the organization as a system, is the sheer complexity that such a view introduces. A model can be helpful in dealing with this complexity. As mentioned in §3.1, such an abstraction of the object of study is typically associated with the epistemology of the functionalist paradigm. Scott (1992) performed a meta-theoretical research study on the subject of organizations from a sociological point of view. In this study, he reviewed roughly 750 books and articles on the field of organization study and related areas, such as anthropology, psychology, and economics. To identify the characteristics of the various research schools, Scott applied a basic model to describe the focal points of the organization as the object of study. He emphasizes that these focal points – or elements – should be studied in an interrelated way. This approach is in line with the General Systems Theory (see §3.2 for more details), where the elements are presumed to 'act' coherently. Scott assumes the development of the study of organizations to be linear. He labeled three major developments, which are successively rational systems, open systems, and natural systems. This presumed linear development is opposite to the assumption of Burrell and Morgan, who identified several coexisting paradigms (see §3.1 for more details). Such a linear development is also typically linked with the functionalist paradigm.

Figure 3-2 represents the model applied by Scott. The gray border represents the boundary between the organization and its environment. How such a boundary can be obtained is merely another focal point of Scott's research, since this is not implied by his model. The organization itself consists of four interrelated elements, which are regarded as equals.

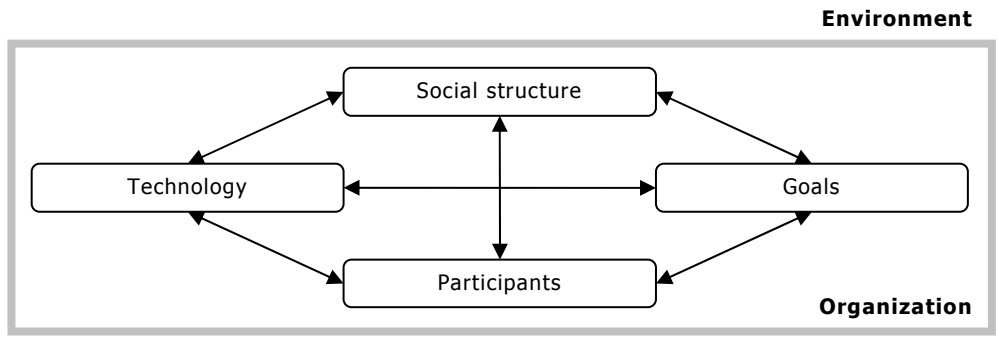

**Figure 3-2: Scott's elements of organization**

The *social structure* can be analytically separated into two interrelated components. The first is the normative structure, which can be regarded as the formal structure of the organization. It acts as a prescription for desired behavior of the organization's participants, and includes values, norms, and roles. This structure is complemented by the behavioral structure. In likely contrast with the prescribed behavior, the behavioral structure describes the actual behavior of participants. For instance, recurrent corridor chat between several participants could indicate a balance of power that is different from the formal hierarchy within an organization. The social structure as a whole especially refers to those relationships that can be characterized as patterned of regularized. *Participants* are those individuals that contribute to the organization in some way or another. In return they receive a variety of incentives, which can include salary or appreciation of one's work. Those who 'own' the organization are likely to have several *goals* they wish the organization to fulfill. These goals can include maximization of profit, or taking care of patients. Finally, the *technology* element represents the means to produce some final product or deliver some kind of service. This element includes equipment, but also knowledge and skills or participants. The latter is a necessary prerequisite to transform the raw materials involved – being physical, informational, or human.

If Scott's model is used to reflect the organizational orientation of BPR, several aspects can be identified. The primary concern of BPR is radical performance improvement of the organization's core processes. Those processes can e.g. be the manufacturing of a product, or a consulting service provided for customers. The *technology* element in Scott's model represents the means to provide these products or services. Therefore the major concern of the BPR approach is technological. Business processes are the dominant conception of this focal point of the study of organizations. The viability of business processes is judged according to their performance and relevance to the organization's strategy. A detailed discussion of Scott's focal points of organization, and the interpretation according to BPR, is presented in Table 3-2.

| Focal point | Interpretation in BPR |
|---|---|
| Social structure | The vertical hierarchy should be overthrown and middle-management eliminated. The normative structure is of primary concern, because it can be designed. The behavioral structure is ignored, or expected to be deduced from the designed structure. |
| Participants | Participants are ignored during the first parts of the change program. They can be convinced from the advantages of this program by using a rationale that is rooted in visionary, strategic remarks[19]. This is in line with a top-down management approach. |
| Goals | The goals of the organization's owners are embedded in a corporate strategy. This strategy is accompanied by a performance program, which enables measurement of |

---

[19] The following quote from Hammer and Champy (1993: 148) is an exemplary illustration: '*Getting people to accept the idea that their work lives – their jobs – will undergo radical change is not a war won in a single battle. It is an educational and communications campaign that runs from reengineering's start to its finish. It is a selling job that begins with the realization that reengineering is required and doesn't wind down until well after the redesigned processes have been put into place.*'

| | |
|---|---|
| | the realization of these goals. IT should play a major role in the strategy. |
| Technology | Business processes are the major focal point of organizations. Benchmarking and performance measurements are rational instruments to support decision-making of the selection and implementation of those processes. Employees have to be reeducated to cope with the new job prescriptions. |
| Environment | The environment is turbulent, because competition is growing stronger due to globalization. In addition, consumers are more demanding. This favors mass customization over mass production. |

**Table 3-2: BPR's view on the focal points of organization**

Clearly the interpretation of the focal points indicates that the model of Scott does not completely reflect BPR's perception of organizations. While Scott stresses the neutrality of the various elements, the strategy and the means to reach its objectives are dominant within BPR. This is reflected in the MIT90s framework of Scott Morton. The description of this framework is as follows (Scott Morton 1991: 18-9).

> An organization can be thought of as comprised of five sets of forces in dynamic equilibrium among themselves even as the organization is subjected to influences from an external environment. In this view, a central task of general management is to ensure that the organization, that is, all five 'forces' (represented by the boxes), moves through time to accomplish the organization's objectives.

Figure 3-3 gives a visual representation of this framework. The central role of management within organizations is eminent. It is the principal part of organizations that ensures the organization responds to changing demands in the external socioeconomic environment. Managers also keep track of technological developments, and decide if such developments are incorporated.

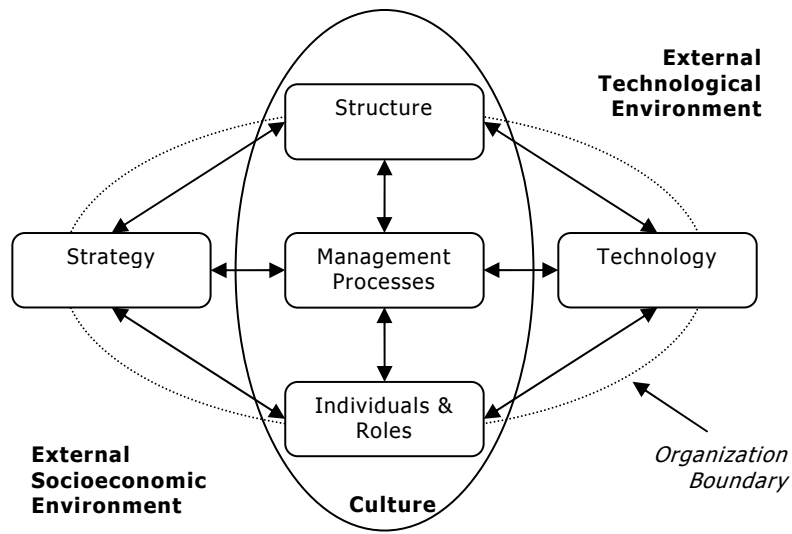

**Figure 3-3: The MIT90s framework (adapted from Scott Morton 1991: 20)**

## 3.4   Conclusions

Sociology is a discipline that studies the collective behavior of individuals. Within sociology many different schools of research exist. They can be differentiated by their philosophy, which is at least composed of the view on the nature of organizations (the ontology) and how one can obtain knowledge about this nature (the epistemology). The dominant perspective within sociology is the so-called functionalist paradigm. Its supporters claim that knowledge – obtained with a rigorous scientific research methodology – about organizations will ultimately lead to a single, coherent theory. The General Systems Theory has had a lot of influence on this paradigm. Within systems the interdependency of elements is of primary concern, because elements behave differently in a collective. If organizations are viewed as a system, the enormous complexity of this system is paramount. Many social theorists therefore have misapplied the systems theory, and used an early analogy with mechanisms or organisms.

Process thinking as illustrated by Business Process Reengineering supports a technocratic management perspective of organization. In such a view, organizations are the means to accomplish the objectives of the organization's owners. These objectives are embedded within a corporate strategy that is realized by business processes. The accompanying technological focal point of organizations views people as resources to enable these processes. The conception of organization according to Business Process Reengineering fits within the aforementioned functionalist system view. This means that elements that cannot be made explicit in such a system are likely ignored. In line with a top-down management approach, the system's conception of the organization's owners will manifest itself as an undisputable reality. While the original envisioning of the system by its founders was directed at a holistic view of social reality, the misapplication of system's theory within BPR resulted in a reductionist perception instead.

## 4 Beyond Process Thinking

The limitations of process thinking are perhaps best illustrated with the high failure rate of Business Process Reengineering projects. The previously identified management orientation of process thinking is reflected using the paradigms of social theory of §3.1 and a functionalist orientation. With the identification of several limitations, some openings to overcome these limitations within the area of Information Systems Development can be presented.

### 4.1 Critical Analysis of Business Process Reengineering

As mentioned in §2.2, the Business Process Reengineering movement had its peak in the early 1990s. BPR largely remained a management practice during the first couple of years, as Barothy, Peterhans et al. (1995) only called for the emergence of BPR as a scientific research field in 1995. The high risk of BPR change programs was already brought up in §2.1. Hammer and Stanton (1995) indicate that 50 to 70 percent of BPR endeavors actually fail. On the demarcation in time between management and science, both founders of the BPR movement acknowledge that their theory has one fatal flaw: it ignores people (Davenport 1995; White 1996). Although there is an immediate appeal to this explanation of the high failure rate of BPR projects, I think it is merely a symptom. If process thinking is seen as the epistemology of BPR to understand organizations, then I suggest that the real cause of BPR's problems lies within its conception of organization.

Chapter 2 discussed Business Process Reengineering from a functionalist perception. The applied post-modern periodization – with use of Scott's model of organization[20] – led to the conclusion that the BPR movement adheres to a technocratic management perspective. In such a view, organizations are the means to realize the goals of its owners. If this outcome is reflected using the social theory of Burrell and Morgan (see §3.1 for an elaboration), two aspects can be identified. The focus on a top-down strategy implies that the conception of the organization's owners is dominant. Instead of accepting a more subjective reality, where discourse between all the participants of the organization – including management and employees – is supported, an objectivist viewpoint of organizational reality is adhered to. Next, BPR is concerned with radical change as opposed to improvement. Literally, reengineering seeks to overcome the bureaucratic organization by getting rid of formal structures, functional work arrangements and vertical hierarchies. Combining these two observations leads to the conclusion that BPR fits within the Radical Structuralist paradigm of Burrell and Morgan.

---

[20] This model is discussed in §3.3.

The supporters of the BPR movement uphold the bureaucracy as the dominant type of organization from the late 18th century and onwards. To reengineer these bureaucratic organizations, their bureaucratic functions are replaced with another mechanism – the business process (Morgan 1997: 22). Paradoxically, this is a reincarnation of the approaches found in organizations in the early decades of the 20th century, while BPR was about the overhaul of such bureaucracies in the first place (Morgan 1997: 386). This perception of organizations can be understood in terms of the machine metaphor of Gareth Morgan (1997: 11-31). If the organization is seen as a machine, then the first step is to set goals and objectives – the strategy. Next one has to organize the means to accomplish this strategy by organizing rationally, efficiently, and clearly. Then the human factor will fall into place automatically, since they are submitted to the designed system. Although BPR tries to eliminate the endless formal specifications of work-breakdown structures – wherein each and every job detail is specified – it is at least questionable if a formal business process specification as a replacement is a real improvement. A fixed set of business processes is more likely to restrain the creativity of people than to encourage it. At best, this hinders the organization's ability to respond to a dynamic environment. At worst, this will reform the organization into the same sluggish machine it was trying to overcome in the first place.

Even from a traditional, functionalist point of view the theory of BPR is flawed. Henry Mintzberg, a renowned researcher within management science, studied several forms of organizations from a structural orientation. In his theory, the only type of organization that is capable to respond to a complex and dynamic environment is the adhocracy[21]. Similarly to BPR's process organization, an adhocratic organization integrates several functions into a single job. Specialists are grouped in functional units for housekeeping purposes only, since they really perform their work in project teams. But contrary to BPR's notion, any form of standardization for coordination should be prevented, because this hinders innovation. This means that 'information and decision process flow flexibly and informally' (Mintzberg 1979: 433). Although this introduces a lot of overhead, it is a necessity if creativity is valued. BPR standardizes communication within a set of business processes. Although the employees are empowered and entitled to make some authorizations, they are ultimately bound to a corporate-wide strategy. This strategy is top-down, while Mintzberg stresses that strategy should be obtained bottom-up and inside-out within the adhocracy. Therefore, BPR cannot uphold its claim that it is a solution to organizations in a dynamic environment. BPR can be helpful in situations where performance improvement is required, but it restrains the organization's agility.

### 4.2   Possible Directions to Overcome Some Limitations of Process Thinking

If process thinking is applied to define the context of Information Systems – a subject that is part of Information Systems Development – a conception of organization similar to that of Business Process Reengineering is used. In such a conception organizations are a system of rational actors of which the behavior can be designed. Actual behavior is a blind spot, since it would be the result of irrational behavior. I therefore suggest that process thinking in isolation will not contribute to more effective Information Systems. I think that the solution of the social problems encountered in Information Systems Development requires a broader perspective. Most and for all, I think that the acceptance of the subjectivity of one's knowledge – opposed to objectivist claims – is a prerequisite to reach such a perspective.

Within the bounds of systems theory, I present several directions that can contribute to the effectiveness of Information Systems Development.
- *Construct the system using a participative discussion between stakeholders.*
  To prevent the manifestation of a system's view that is constructed by a limited group of designers, the Soft Systems Methodology of Peter Checkland – discussed in §3.2 – could be helpful. Although several functionalist approaches – such as aspect models and rational discussions – are used in this methodology, at least all the stakeholders are aware of their limited perception of the situation at hand. As long as the group of participants is a good representation of all stakeholders, their final system definition could be an acceptable substitute for the problem situation.
- *Apply a more balanced view of the organizational system.*
To obtain a more balanced system's view of the organizational context of Information Systems, the organizational model of Scott – discussed in §3.3 – could be used. While process thinking is largely concerned with the focal point of technology, Scott stresses that the other elements – social structure, goals, and participants – and the environment of the organization should receive equal attention.

---

[21] The 'simple structure' – another form of organization that could be capable of the required flexibility – is dominated by a single owner, which is only appropriate for small companies. Such a company would depend on the owner's ability to adapt to continuous change.

### 4.3   Conclusions

The BPR orientation does not ignore people, but views them from a technocratic management perspective. In this perspective, people in organizations are simply the means to accomplish an objective set out by the owners of these organizations. People are treated as tools that must be controlled and aligned in an efficient organizational machine. This can be accomplished by training and education. In this respect, behavior of the organization's participants is designed within a normative structure, while the actual behavior of those participants is completely ignored. If this prescribed behavior is a starting point, expecting people to be creative and be able to react to changes in their environment is a fallacy. Being creative is not simply an objective that can be set, but is only enabled and stimulated in a proper organizational setting. Therefore organizations that are subject to BPR's perspective will not be able to cope with the requirements of a turbulent and dynamic environment, since such environments require creativity and agility of organizations. Business Process Reengineering falls in the same trap of the mechanistic organizations it was trying to counter. The high failure rates of BPR projects do not occur because their practitioners misinterpreted and misapplied the theory, but are a result of the theory itself. There is no quick fix to BPR, since the entire foundation of the movement is flawed. To overcome its limitations, the entire concept of BPR must be abandoned. As an alternative, an epistemology rooted in subjectivism is likely a better direction for research to support Information Systems Development.

Process thinking as illustrated by BPR is an extreme example. In reality, the application of such an orientation will not be that extreme, but likely converges with other orientations and practices. Nevertheless, it does point to what process thinking could lead to if its application remains unquestioned. Within the unification of the General Systems Theory several directions can prove helpful to overcome some limitations of process thinking. Although a system's view inevitably will manifest itself as a substitute for organizational reality – and therefore limits the perceptive ability of its studiers – a somewhat more subjective orientation can be applied by means of a participative discussion between stakeholders. In addition, if the participants give equal attention to different focal points of the organization, such a system could be an acceptable reproduction of organization. However, these directions will slow down the process of Information Systems Development. Within an organizational setting that focuses on quick wins, it might prove helpful to limit the scope of IT projects. If the directions are applied successfully in some small projects, it might convince critics of their feasibility.

## 5   Conclusions

The implementation of business processes has a profound impact on the structure of organizations. In addition, the supporters of the process movement advocate the strategic importance of Information Technology. Therefore the structural complexity of Information Systems increases. However, this is largely a technical problem that does not require a different perspective of those who study it.

The real cause of the high failure rates of IT projects is not technical, but lies within their social context. If stakeholders are confronted rather than consulted, resistance to Information Systems increases and will continue to do so in the future. The dominance of process thinking within Information Systems Development contributes to this development. Process thinking conceives organizations as mechanical systems, and therefore expects people to fall into place automatically. This perspective could be a result of the mindset of IT scientists and practitioners, since Information Systems are a technical construction that can be designed completely. But to treat organizations as similar systems leads to many problems, of which resistance to new Information Systems is just an effect that comes to the surface.

Therefore, process thinking is limited in the way it views organizations. Within the General Systems Theory several directions are possible to overcome some of the limitations of process thinking. But inevitably a system's view manifests itself as a fallacious substitute for organizational reality, because it makes elements explicit and is bound to rational behavior. The perception of social reality should be continuous and dynamic, and not linearly directed at a fixed endpoint. Organizations and their context are fluctuating too, and therefore any understanding is finite. The required reflexive perception is bound to subjectivity, which means a single universe of discourse is unattainable. Therefore the true challenge of requirements engineers is not to reach an acceptable – or even optimal – design, but to reach such a design while dealing with the conflicts that a subjective universe entails.

This research is certainly limited. I have based the conclusion on rational arguments from a single point of view; that of my own. In addition, I have solely performed a theoretical study and did not investigate any cases in real life. This paper aims to inspire fellow IT researchers and practitioners to at least question the applicability of process thinking.

## Acknowledgements

Thanks to Rogier Krieger for critical reading and discussion of the manuscript, to Sjoerd van Tuinen for critical reading and his comments on social theory, and to Jan Dietz and Hans Mulder for their comments on the application of business processes.

22    *M.J. Dumay / Business Processes: The Theoretical Impact of Process Thinking on ISD*Petrovic, O. (1995). *On the Necessity of an Iterative Design of Business Strategy, Business Organization and Information Technology*. Proceedings of the 28th Annual Hawaii International Conference on System Sciences.

Riesewijk, B. and J. Warmerdam (1988). *Het slagen en falen van automatiseringsprojecten; een onderzoek naar de sociaal-organisatorische implicaties van automatisering voor gebruikersorganisaties en computer service bedrijven*. Nijmegen, Instituut voor Toegepaste Sociale Wetenschappen.

Schwartz, B., B. W. Hwang, et al. (1995). *A workplan for business process reengineering and a challenge for information science and technology*. Proceedings CSC'95: 23rd Annual ACM Computer Science Conference. 28 Feb. 2 March 1995 Nashville, TN, USA, ACM, New York, NY, USA.

Scott Morton, M. S. (1991). *The corporation of the 1990s; information technology and organizational transformation*. Oxford, Oxford University Press.

Scott, W. R. (1992). *Organizations; rational, natural, and open systems*. 3rd edition Englewood Cliffs, Prentice-Hall.

Silvestro, R. and C. Westley (2002). "Challenging the paradigm of the process enterprise: a case-study analysis of BPR implementation." *Omega* 30(3): 215-25.

Simon, H. A. (1976). *Administrative Behavior; a study of decision-making processes in administrative organization*. 3rd edition New York, Free Press.

Smit, K. (2003). *Presentation slides of ae4-490 Maintenance Engineering*, Faculty of Aerospace Engineering, Delft University of Technology.

The Standish Group International (2003). *Latest Standish Group CHAOS Report Shows Project Success Rates Have Improved by 50%*. The Standish Group International. Web version: http://www.standishgroup.com/press/article.php?id=2.

White, J. B. (1996). *'Next Big Thing': Re-Engineering Gurus Take Steps to Remodel Their Stalling Vehicles*. Wall Street Journal. New York**:** A1, A6.